\documentclass[twocolumn,final,aps,prl,showpacs,superscriptaddress,amsmath,amssymb,amsfonts,floatfix,longbibliography]{revtex4-2}

\usepackage{graphicx}
\usepackage{booktabs}
\usepackage{multirow}
\usepackage{makecell} 
\usepackage{array}
\usepackage{url}
\usepackage{amsmath}   
\usepackage{siunitx}  
\usepackage{ulem}

\usepackage[colorlinks,breaklinks,bookmarks=true,citecolor=blue,linkcolor=blue,urlcolor=blue]{hyperref}
\usepackage{color}
\usepackage{tabularx}
\usepackage{sidecap}
\usepackage{soul}
\usepackage{float}
\usepackage{color}

\usepackage{orcidlink}
\usepackage[colorlinks,breaklinks,bookmarks=true,citecolor=blue,linkcolor=blue,urlcolor=blue]{hyperref}

\begin{document}

\author{Chao Deng\,\orcidlink{0009-0003-8433-2909}}
\affiliation{School of Physics, Northwest University, Xi'an 710127, China}

\author{Motoharu Kitatani\,\orcidlink{0000-0003-0746-6455
}}
\affiliation{Department of Material Science, University of Hyogo, Ako, Hyogo 678-1297, Japan}
\affiliation{Graduate School of Engineering, Kyushu Institute of Technology, Kitakyushu 804-8550, Japan}

\author{Guiwen Jiang}
\affiliation{School of Physical Science and Technology, Soochow University, Suzhou 215006, China}

\author{Siqi Guo\,\orcidlink{0000-0002-5534-7904}}
\affiliation{School of Physics, Northwest University, Xi'an 710127, China}

\author{Niklas Witt\,\orcidlink{0000-0002-2607-4986}}
\affiliation{Institut f{\"u}r Theoretische Physik und Astrophysik and W{\"u}rzburg-Dresden Cluster of Excellence ctd.qmat, Universit{\"a}t W{\"u}rzburg, 97074 W{\"u}rzburg, Germany}

\author{Ao Zhang\,\orcidlink{0000-0001-9786-3942}}
\affiliation{School of Physics, Northwest University, Xi'an 710127, China}
\affiliation{Department of Physics, Southeast University, Nanjing 211189, China}

\author{Wenfeng Wu\,\orcidlink{0000-0002-6575-5813}}
\affiliation{Key Laboratory of Materials Physics, Institute of Solid State Physics, HFIPS, Chinese Academy of Sciences, Hefei 230031, China}
\affiliation{Science Island Branch of Graduate School, University of Science and Technology of China, Hefei 230026, China}
\affiliation{Institute of Solid State Physics, TU Wien, 1040 Vienna, Austria}

\author{Mi Jiang\,\orcidlink{0000-0002-9500-202X}}
\affiliation{School of Physical Science and Technology, Soochow University, Suzhou 215006, China}

\author{Karsten Held\,\orcidlink{0000-0001-5984-8549}}
\affiliation{Institute of Solid State Physics, TU Wien, 1040 Vienna, Austria}

\author{Liang Si\,\orcidlink{0000-0003-4709-6882}}
\email{siliang@nwu.edu.cn}
\affiliation{School of Physics, Northwest University, Xi'an 710127, China}
\affiliation{Shaanxi Key Laboratory for Theoretical Physics Frontiers, Xi'an 710127, China}
\affiliation{Fundamental Discipline Research Center for  Quantum Science and technology of Shaanxi Province, Xi'an 710127, China}
\affiliation{Institute of Solid State Physics, TU Wien, 1040 Vienna, Austria}

\title{Heterostructuring as Gateway to Electron Doping of Nickelate Superconductors}

\begin{abstract}
Despite enormous expenditures in the research field,  the electron-doped side of nickelate superconductors remains uncharted territory. Substituting the trivalent rare-earth cations  by a tetravalent one hitherto failed. Here, we demonstrate by first-principles calculations a disorder-free route to electron dope  Ruddlesden–Popper nickelates. When intercalating wide-band-gap insulating layers such as La$X$O$_3$ ($X$=Al, Ga, Sc) into La$_2$NiO$_4$, the extra (LaO)$^+$ layers act as electron donors, releasing carriers into the Ni-3$d$ orbitals. This electron doping puts La$_2$NiO$_4$:La$_2$AlO$_4$  naturally in the optimal region for  $d_{x^2-y^2}$-wave superconductivity  with $T_{\mathrm{c}}$ exceeding 50\,K.
The same concept also allows us to electron dope La$_3$Ni$_2$O$_7$, the  superconductor in the limelight.
\end{abstract}

\maketitle

\renewcommand{\emph}{\textit}

\emph{Introduction---}
A 
central motif for 
unconventional superconductors such as cuprates \cite{bednorz1986possible,anderson1997theory,RevModPhys.72.969,RevModPhys.75.473}  and   nickelates \cite{li2019superconductivity,zeng2022superconductivity,sun2023signatures,zhao2025pressure,PhysRevX.14.011040}
is doping. This doping is required to suppress the
antiferromagnetic insulating state of the parent compound
with superconductivity emerging in a narrow window 
of doping next to this antiferromagnetic ordering.
In cuprates such as La$_2$CuO$_4$ \cite{bednorz1986possible,anderson1987resonating}, superconductivity emerges upon hole doping via alkaline-earth substitution such as replacing La$^{3+}$ by Sr$^{2+}$ or Ca$^{2+}$. This drives the Cu$^{2+}$ (3$d^9$) state toward a hole-doped, strongly correlated superconducting regime \cite{PhysRevB.37.3759}.
Similarly, the electron-doped side had been mapped out for cuprates by e.g.\
substituting Nd$^{3+}$ with Ce$^{4+}$  in
Nd$_2$CuO$_4$ \cite{Takagi2989,RevModPhys.82.2421}; cf.\
\cite{j2006magnetic,lee2014asymmetry,abbamonte2002structural}.

For nickelate superconductors,  such as infinite-layer (IL) nickelates $R$NiO$_2$  \cite{li2019superconductivity,zeng2022superconductivity}
and  Ruddlesden-Popper (RP) nickelates  $R_{n+1}$Ni$_n$O$_{3n+1}$ \cite{sun2023signatures,zhao2025pressure,PhysRevX.14.011040,ko2025signatures,zhou2025ambient}, 
hole doping through substituting the rare-earth cation $R^{3+}$ by Sr$^{2+}$ or Ca$^{2+}$
is possible \cite{PhysRevLett.125.027001,zeng2022superconductivity,osada2021nickelate,wang2025recent,ko2025signatures,zhou2025ambient}, and laid bare the enormous potential of nickelates. 
While hole doping of nickelates is straightforward, the lack of a robust route to electron doping has hindered a complete mapping of their phase diagram. 
This difficulty arises because substitutional doping requires higher-valence cations, i.e., tetravalent $A$-site ions. For LaNiO$_3$ \cite{puphal2023phase} with Ni$^{3+}$ just as in cuprates \cite{RevModPhys.82.2421}, this is possible by substituting La with Ce$^{4+}$. However, because the fourth electron ionic energy of Ce$^{3+}$ is higher than the second ionization energy of Ni$^{1+}$ for RP nickelates this route is not possible.

\begin{figure}[tb]
\centering
\includegraphics[width=0.9\linewidth]{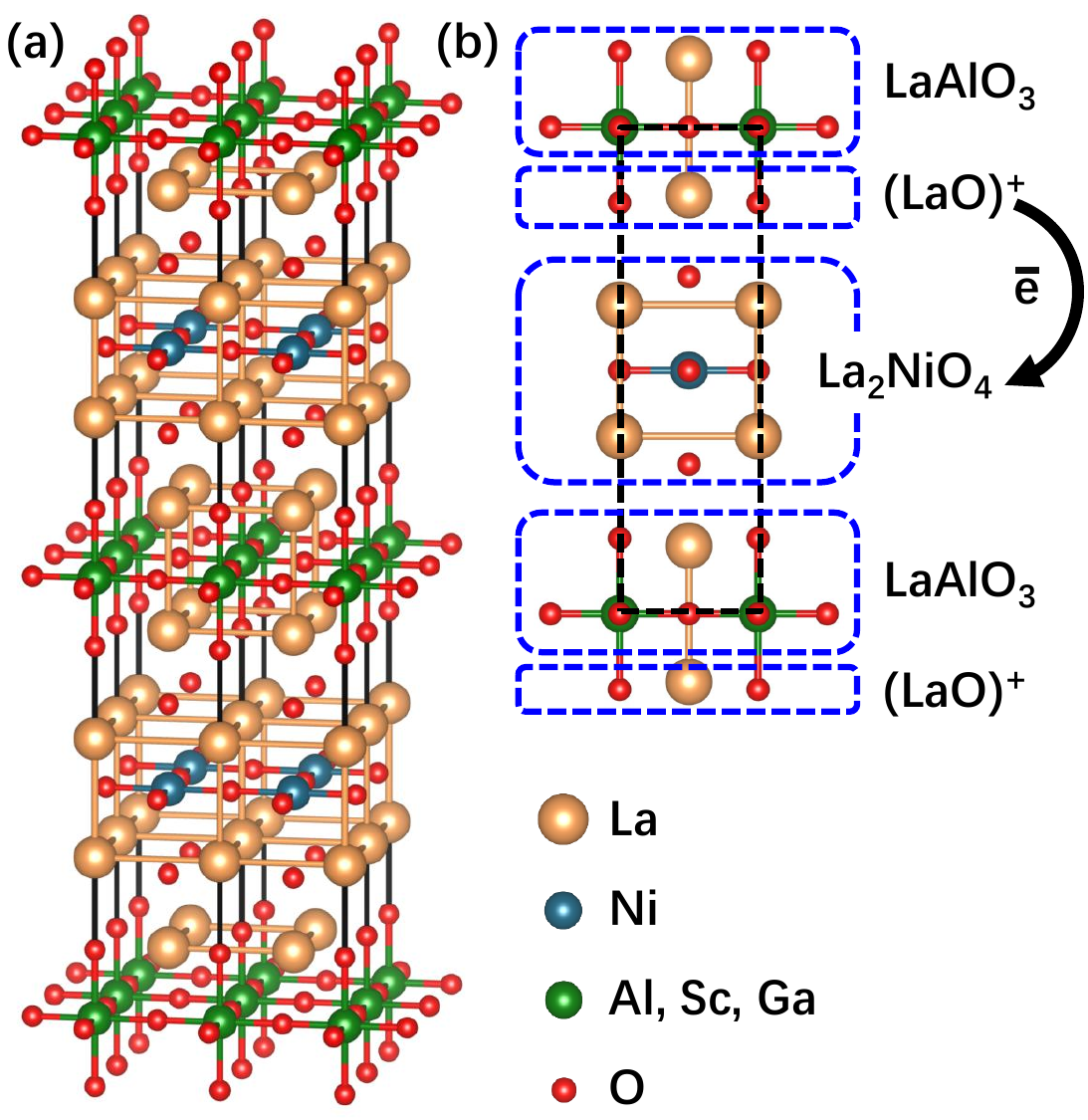}
\caption{(a) Structural schematic of La$_2$NiO$_4$:La$_2$$X$O$_4$ ($X$=Al, Sc, or Ga); (b) Mechanism of intrinsic electron doping in La$_2$NiO$_4$ realized by inserting additional (LaO)$^+$ layers. The black dashed line in (b) denotes the unit cell used for the DFT and DMFT calculations shown in Figs.~\ref{Fig2_DFT+DMFT-band}(b,d).}
\label{Fig1_structure}
\end{figure}

Here, we propose a conceptually simple but feasible solution, see Fig.~\ref{Fig1_structure}: achieve electron doping via wide-band-gap insulator intercalation in RP lanthanum oxides. The basic principle exploits the atomic and elementary versatility of RP phases, which consist of perovskite blocks separated by rocksalt layers. By inserting wide-gap insulating La$X$O$_3$ blocks ($X$ = Al, Ga, Sc) \cite{rizwan2019review,lybye2000conductivity}, which naturally form La$_2$$X$O$_4$ units in RP phase structures, the lattice introduces excess (LaO)$^+$ layers. These act as effective electron donors, releasing electrons into adjacent transition-metal oxide planes. Unlike conventional chemical doping, this strategy does not introduce chemical disorder at $A$ and $B$ sites; hence, it provides a defect- and disorder-free  symmetry-preserving route to electron doping. This is very advantageous for stabilizing \emph{fragile} emergent correlated states. We investigate this concept  through first-principles and many-body calculations for La$_2$NiO$_4$:La$_2$$X$O$_4$ heterostructures.

Our density-functional theory (DFT) \cite{PhysRev.136.B864,PhysRev.140.A1133} and dynamical mean-field theory (DMFT) \cite{PhysRevLett.62.324,RevModPhys.68.13} calculations confirm significant charge transfer from the La(Al,Ga,Sc)O$_3$-derived blocks into NiO$_2$ layers, driving Ni toward 3$d^9$ configurations in La$_2$NiO$_4$ units and electron doping La$_3$Ni$_2$O$_7$. Furthermore, many-body  calculations identify $d_{x^2-y^2}$-wave superconductivity (above 50\,K) in La$_2$NiO$_4$:La$_2$AlO$_4$  even at ambient pressure. These results establish wide-band-gap intercalation as a promising strategy to access electron-doped nickelate. The proposed strategy may be extended to lanthanum manganites, cobaltates, or ruthenates, where the research on electron-doping effects still faces challenges, and even cuprates.

\emph{Method---}The calculations in this study were primarily performed using DFT \cite{PhysRev.136.B864,PhysRev.140.A1133,PhysRevB.50.17953,kresse1996efficiency,PhysRevB.54.11169,blaha2001wien2k,Schwarz2002,PhysRevLett.77.3865} for structural optimization and band structure analysis. The constrained random phase approximation \cite{PhysRevB.77.085122,PhysRevMaterials.9.015001,PhysRevLett.124.166402} is employed for determining interaction parameters, and DMFT \cite{PhysRevLett.62.324,RevModPhys.68.13,PhysRev.52.191,RevModPhys.84.1419,mostofi2008wannier90,kunevs2010wien2wannier,PhysRevB.52.R5467,PhysRevB.48.16929,RevModPhys.83.349,PhysRevB.86.155158,wallerberger2019w2dynamics,PhysRevB.44.6011,PhysRevB.57.10287,kaufmann2023ana_cont} to account for local correlation effects. Additionally, the dynamical vertex approximation (D$\Gamma$A) \cite{galler2019abinitiodgammaa,held2008dynamical,RevModPhys.90.025003,PhysRevB.99.041115,Toschi2007}, fluctuation exchange (FLEX) method \cite{Kitatani2022,PhysRevLett.62.961,PhysRevB.43.8044,PhysRevB.103.205148,10.21468/SciPostPhys.15.5.197,PhysRevB.101.035144,PhysRevB.96.035147,wallerberger2023sparse}, and dynamical cluster approximation (DCA) \cite{hettler1998nonlocal,gull2008continuous,maier2006structure,maier2006pairing,maier2005quantum} were employed to study correlations beyond DMFT and to estimate the critical superconducting temperatures $T_c$. 
(Self-consistent) phonon calculations (SCPH) \cite{togo2015first,alamodePRL,alamodePRB} and first-principles molecular dynamics (AIMD) \cite{PhysRevB.48.13115} simulations using on-the-fly machine-learned force fields \cite{PhysRevLett.122.225701,PhysRevB.100.014105} were carried out to assess the structural stability. Additional results, computational details, and parameter settings are provided in Supplemental Materials (SM) Secs.~I-IX \cite{SM}.

\emph{Structure---}We focus on RP-phase oxides $A_{x+1}B_x$O$_{3x+1}$ with $x$=1, i.e., $A_2$$B$O$_4$. Many perovskite transition-metal oxides have been synthesized in this RP phase; prominent examples include Sr$_2$RuO$_4$ \cite{luke1998time}, La$_2$CuO$_4$, and Sr$_2$TiO$_4$ \cite{jacob2011thermodynamic}. These compounds feature $A_2$$B$O$_4$ layers stacked on top of each other. Often they are synthesized by molecular beam epitaxy or pulsed laser deposition.
For $x$=$\infty$, on the other hand, combinations of trivalent rare-earth cations (e.g., La$^{3+}$) with trivalent elements from the III\,A or III\,B group form wide-band-gap perovskite insulators such as LaAlO$_3$ (3.6\,eV), LaGaO$_3$ (2.9\,eV), and LaScO$_3$ (2.9\,eV).
For $AB$O$_3$ compounds, heterostructures or interfaces can be engineered when two materials share similar crystal structures and lattice constants. Consequently, when growing the $A$$_2$$B$O$_4$---whose unit blocks contain an $AB$O$_3$ monolayer---it becomes possible to insert additional $AB$O$_3$ layers of varying type or thickness. Examples include Sr$_2$TiO$_4$/SrTiO$_3$ ($x$=1+$\infty$) interfaces \cite{li2021interfacial} and the recently reported La$_2$NiO$_4$/La$_3$Ni$_2$O$_7$ heterostructure ($x$=1+2) \cite{shi2025pressure,PhysRevMaterials.8.053401}.

Following this reasoning, replacing every other unit block of  La$_2$NiO$_4$  by La$_2$$X$O$_4$ yields structural analogs where La$_2$$X$O$_4$ (La$_3$$X_2$O$_7$) can be viewed as one (or two) unit cells of La$X$O$_3$ plus an extra (LaO)$^+$ layer. Here, La and O ions are expected to retain their 3$+$ and 2$-$ valence states, respectively. This introduces an extra electron into the La$_2$NiO$_4$ layer, provided that La and $X$ (Al, Ga, and Sc) are more ionic than Ni, as illustrated in Fig.~\ref{Fig1_structure}(b). 
In Supplemental Material Sec.~IX and X \cite{SM}, we employed DFT, SCPH and AIMD methods to perform phonon spectra, finite-temperature MD simulations, variable-composition structural searches and constructed convex hulls. The results demonstrate that the La$_2$NiO$_4$:La$_2$AlO$_4$ heterostructure remains dynamically,  chemically and thermodynamically stable at even 300\,K.

\begin{figure*}[t]
    \centering    \includegraphics[width=0.95\linewidth]{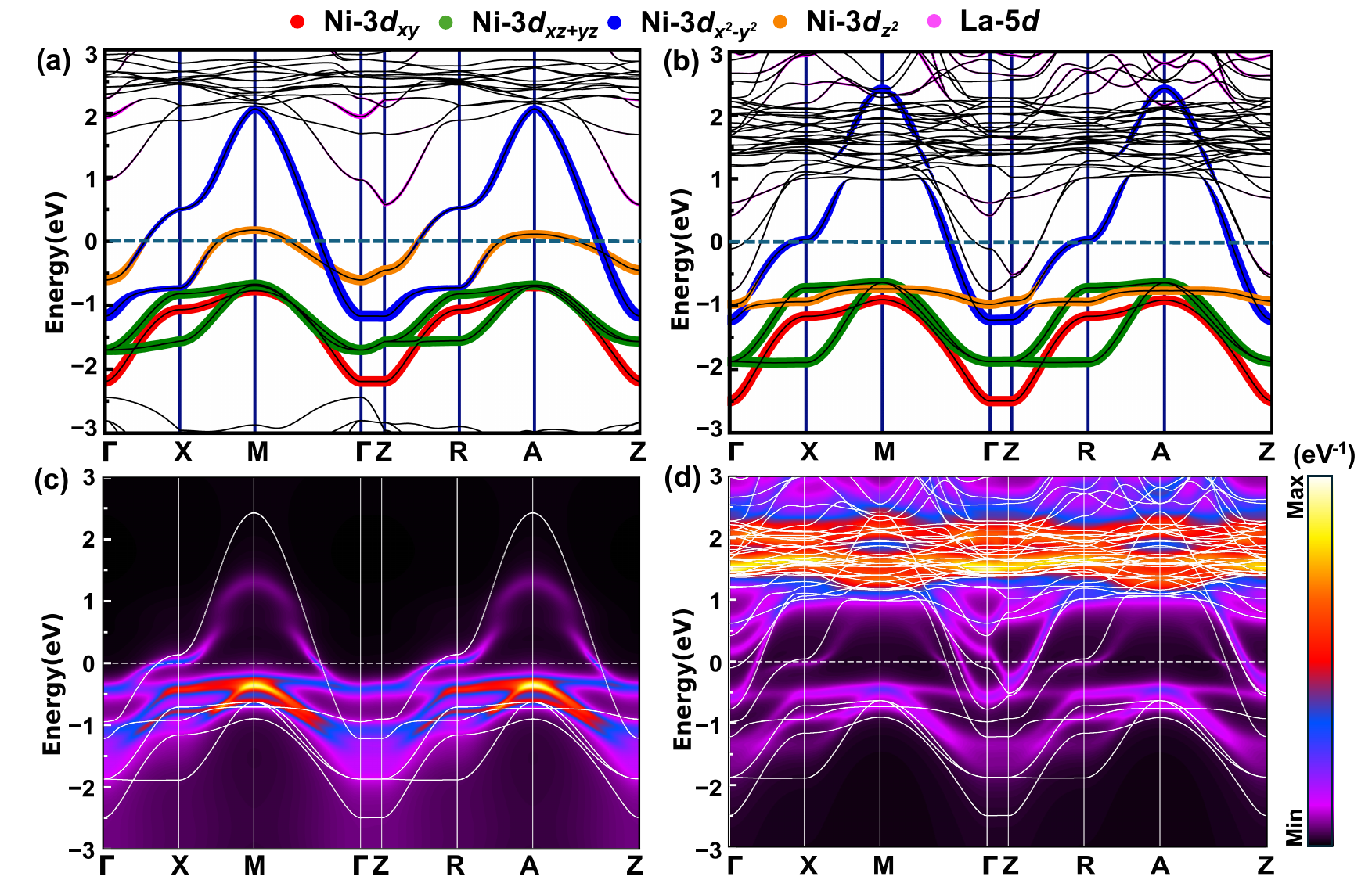}
    \caption{Paramagnetic DFT band structure  and orbital characters of (a) La$_2$NiO$_4$ and (b) La$_2$NiO$_4$:La$_2$AlO$_4$. Multiband DMFT $k$-resolved spectral functions $A$($k,\omega$) of La$_2$NiO$_4$:La$_2$AlO$_4$ with (c) Ni-3$d$-only model and (d) La-4$f$-5$d$+Ni-3$d$ model.}
    \label{Fig2_DFT+DMFT-band}
\end{figure*}

\begin{figure*}[ht]
\centering
\includegraphics[width=0.95\linewidth]{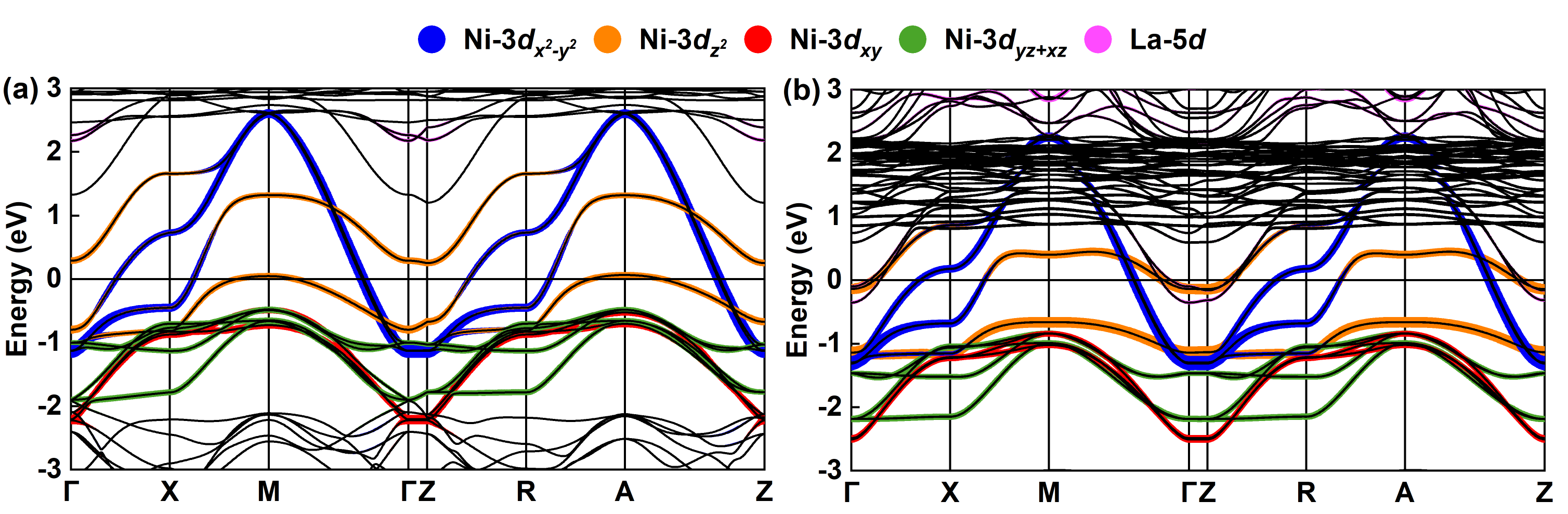}
\caption{Paramagnetic DFT band structure  and orbital characters of (a) La$_3$Ni$_2$O$_7$ and (b) La$_3$Ni$_2$O$_7$:La$_3$Al$_2$O$_7$.}
\label{Fig3_327}
\end{figure*}

\emph{DFT bands---} Let us start with the DFT band structure [Fig.~\ref{Fig2_DFT+DMFT-band}(a,b)]  and orbital occupations (Table~\ref{Tab1}) of La$_2$NiO$_4$ with and without the intercalation of La$_2$$X$O$_4$. For (undoped) La$_2$NiO$_4$, Ni ions adopt a valence state of 2+ (3$d^8$), with fully occupied Ni-$t_{2g}$ and half-filled Ni-$e_g$ orbitals; see Table~\ref{Tab1}. Because of the elongated octahedral field in the $z$-direction and, thus, more distant (negatively charged) O-2$p_z$ orbitals, the Ni $d_{z^2}$ orbital is lower in energy than the $d_{x^2-y^2}$ orbital; see Fig.~\ref{Fig2_DFT+DMFT-band}(a). Consistent with this, Table~\ref{Tab1} shows an occupancy of 1.46 electron in $d_{z^2}$ and 0.56 electron in $d_{x^2-y^2}$ in DFT. When La$_2$AlO$_4$ is intercalated [Fig.~\ref{Fig2_DFT+DMFT-band}(b) and Table~\ref{Tab1}], the additional (LaO)$^+$ layers release roughly one electron to Ni-3$d_{z^2}$, leading to the occupation of $n$$\sim$1.91 (1.93) in DFT (DMFT), while the $d_{x^2-y^2}$ occupation increases slightly to 0.65 (0.87) in DFT (DMFT). This corresponds to approximately one electron doping per Ni as in the sketch Fig.~\ref{Fig1_structure}{(b)}.

Similar electron-doping effects are observed for La$_2$NiO$_4$:La$_2$GaO$_4$ and La$_2$NiO$_4$:La$_2$ScO$_4$, the corresponding Table~SI and Fig.~S4 can be found in Supplemental Material \cite{SM} Secs.~XI and XII. However, as the ionic size increases from Al via Ga to Sc, the in-plane lattice expands (see Table~\ref{Tab1}), inducing a compressive strain in the $z$ direction. This lowers the energy of the $d_{x^2-y^2}$ orbital, decreases its occupation, and increases the occupation of  the $d_{z^2}$ orbital. Importantly, in La$_2$NiO$_4$:La$_2$AlO$_4$, the Fermi surface exhibits quasisingle $d_{x^2-y^2}$ band character, with nearly full $d_{z^2}$ occupancy and half-filled $d_{x^2-y^2}$, closely resembling cuprates such as La$_2$CuO$_4$ \cite{PhysRevB.98.125140,PhysRevB.49.14211} and CaCuO$_2$ \cite{PhysRevX.10.021061} and hydrogenated La$_2$NiO$_4$ \cite{yd8w-frs8}. At the $\Gamma$=(0,0,0) and $Z$=(0,0,$\pi$) momenta, pocket bands of predominately La-5$d$ character emerge but do not overlap with Ni-$e_g$ bands. 
These pockets lead to the La occupation of 0.16 (0.10) electrons/La in DFT (DMFT) and self-dope the Ni-3$d_{x^2-y^2}$ band,  analogous to IL nickelates \cite{PhysRevLett.125.077003,PhysRevB.100.205138}. Moreover, spin-polarized DFT+$U$ calculations reveal that the proposed compounds exhibit (in-plane) $G$-type antiferromagnetism, analogous to that observed in 2D cuprates (see Supplemental Material Sec.~XI \cite{SM}). 

In addition to the DFT bands, the charge transfer effect induced by the heterostructure is further confirmed by DFT density of states (DOS), Mulliken population and Bader charge analysis (see Supplemental Material Sec.~XIII \cite{SM}).

The mechanism of electron doping can be extended to other RP phase oxides, such as the recently discovered superconductor La$_3$Ni$_2$O$_7$ \cite{sun2023signatures,zhao2025pressure,PhysRevX.14.011040,ko2025signatures,zhou2025ambient}. For La$_3$Ni$_2$O$_7$, we constructed an La$_3$Ni$_2$O$_7$:La$_3$Al$_2$O$_7$ superlattice.
Figure~\ref{Fig3_327}(a,b) 
shows the DFT bands with a pronounced downward shift of the Ni-$e_g$ orbitals, corresponding to an electron-doping of the low-energy Ni orbitals.
This demonstrates the versatility of our electron doping approach and its applicability to diverse RP nickelates and other oxides.

Unlike conventional approaches that rely on the potential difference between $B$ and $B'$ cations at $AB$O$_3$/$AB'$O$_3$ heterostructure to induce charge transfer and self-doping \cite{LaNiO3LaFeO3SciAdv,LaTiO3ROLaNiO3JAP,LaTiO3nLaVO3nPRB}, the gateway we propose avoids introducing additional bonding associated with electron or hole doping, which typically alters the electronic and magnetic structures.

\begin{table*}[t!]
\centering
\caption{Occupation of Ni-3$d$ and La-5$d$ orbitals, out-of-plane Ni-O bond, and in-plane lattice constants (in unit of \AA).}
\label{Tab1}
\vspace{0.38cm} 
\renewcommand{\arraystretch}{1.0}
\setlength{\tabcolsep}{5pt}
\begin{tabular}{c c c c c c c c c}
\hline\hline
\multirow{3}{*}{\makecell{Materials\\Wannier\\Projection\\Basis}} & 
\multirow{3}{*}{\makecell{Method}} & 
\multicolumn{5}{c}{Orbital Occupancy} &
\multicolumn{2}{c}{Structural Parameters (\AA)} \\
\cmidrule(lr){3-7} \cmidrule(lr){8-9}
 & & \makecell{$d_{xy}$} & \makecell{$d_{xz}/d_{yz}$} & \makecell{$d_{x^2-y^2}$} & \makecell{$d_{z^2}$} & \makecell{Average-\\La-5$d$} & 
\makecell{Out-of-\\plane\\Ni-O} & 
\makecell{In-plane\\Lattice\\Constant} \\
\hline
\multirow{2}{*}{\makecell{La$_2$NiO$_4$\\(Ni-3$d$ only)}} 
& DFT & 1.98 & 2.00 & 0.56 & 1.46 & -- & \multirow{2}{*}{2.37} & \multirow{2}{*}{3.80} \\
& DMFT & 2.00 & 2.00 & 1.00 & 1.00 & -- & & \\
\addlinespace[2pt]
\multirow{2}{*}{\makecell{La$_2$NiO$_4$:La$_2$AlO$_4$\\(La-5$d$+Ni-3$d$)}} 
& DFT & 1.95 & 1.92 & 0.65 & 1.91 & 0.16 & \multirow{2}{*}{2.56} & \multirow{2}{*}{3.76} \\
& DMFT & 1.95 & 1.93 & 0.87 & 1.93 & 0.10 & & \\
\addlinespace[2pt]
\multirow{2}{*}{\makecell{La$_2$NiO$_4$:La$_2$GaO$_4$\\(La-5$d$+Ni-3$d$)}} 
& DFT & 1.97 & 1.96 & 0.67 & 1.90 & 0.13 & \multirow{2}{*}{2.50} & \multirow{2}{*}{3.80} \\
& DMFT & 1.98 & 1.96 & 0.90 & 1.93 & 0.06 & & \\
\addlinespace[2pt]
\multirow{2}{*}{\makecell{La$_2$NiO$_4$:La$_2$ScO$_4$\\(La-5$d$+Ni-3$d$)}} 
& DFT & 1.94 & 1.92 & 0.76 & 1.81 & 0.16 & \multirow{2}{*}{2.44} & \multirow{2}{*}{3.94} \\
& DMFT & 1.96 & 1.94 & 1.01 & 1.71 & 0.11 & & \\
\hline\hline
\end{tabular}
\end{table*}

\emph{Correlations---}Next we turn to the effect of correlations, starting with
the strong local correlations that can be captured by DMFT, the corresponding computational details are provided in Supplemental Material Sec.~III \cite{SM}. As theoretically \cite{kitatani2020nickelate} and experimentally \cite{ding2024cuprate,sun2025electronic} established for both infinite- and finite-layer nickelates as well as for cuprates, quasi-2D superconductivity is primarily governed by the nearly half-filled Fermi surface derived from Ni-3$d_{x^2-y^2}$ single-band. The slightly occupied pocket(s) at $\Gamma$ and $A$ points are considered to play, as in nickelates, a minor role of an electron reservoir (hole donor) and not to be decisive for superconductivity.

\begin{figure}[t!]
\centering
\includegraphics[width=1.0\linewidth]{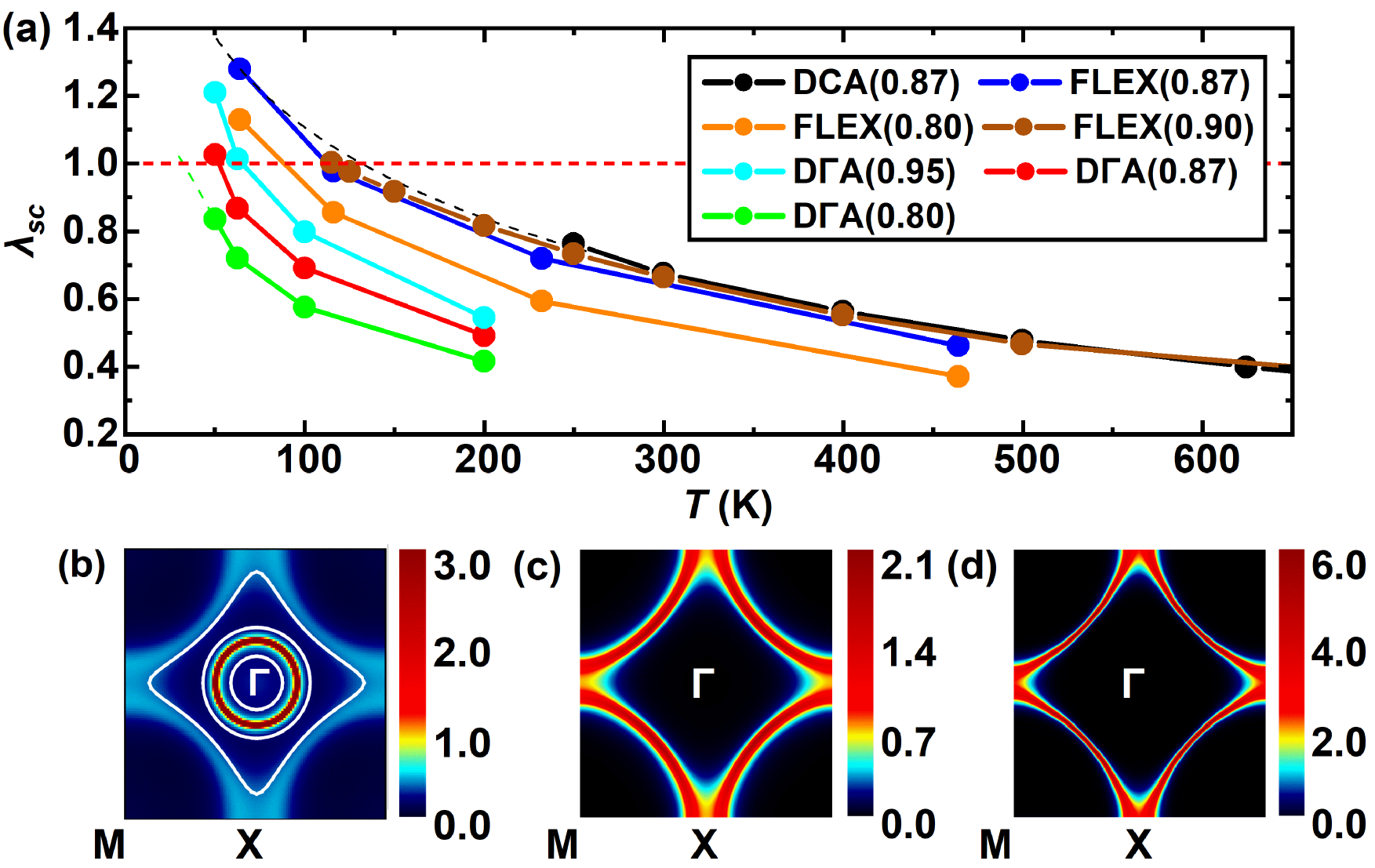}
\caption{(a) Leading $d$-wave superconducting $\lambda_{\mathrm{SC}}$ eigenvalue as a function of temperature using fitting form of $\lambda_{\mathrm{SC}}\approx a- b \ln(T)$ as in previous publications \cite{kitatani2020nickelate,PhysRevLett.130.166002,yd8w-frs8}; Fermi surface at $k_z$=0 computed by (b) DFT (white lines) and DMFT (colorful spectra), (c) FLEX, and (d) D$\Gamma$A. The FLEX and D$\Gamma$A Fermi surface are computed by the momentum dependence of the imaginary part of the Green function [$-\mathrm{Im}\, G(k,\omega_0)/\pi$] at lowest Matsubara frequency $\omega_0$=$\pi k_{\mathrm{B}}T$, $T$=53\,K and $n$=0.87. 
Note that the DFT Fermi surfaces in (b) displays additional $\Gamma$-pockets that are taken into account in D$\Gamma$A and FLEX only through a reduced filling.}
\label{Fig4_Tc}
\end{figure}

We perform DFT+DMFT calculations for two different low-energy models:
including  all  Ni-3$d$ orbitals (only) [Fig.~\ref{Fig2_DFT+DMFT-band}(c)]
and a La-4$f$+La-5$d$+Ni-3$d$ model which also includes the La-4$f$ and -5$d$ orbitals
[Fig.~\ref{Fig2_DFT+DMFT-band}(d)]
\footnote{The strong overlap between La-4$f$ and -5$d$ orbitals around 2\,eV complicates the construction of La-5$d$+Ni-3$d$ Wannier functions, as neglecting La-4$f$ states requires a broad energy window and more DFT Bloch bands, hindering high-quality of the projection}. In the Ni-3$d$ (only) case, both $t_{2g}$ and $d_{z^2}$ bands shift upward by $\sim$1\,eV, while the Fermi surface remains dominated by the $d_{x^2-y^2}$ orbital. The La-4$f$-5$d$+Ni-3$d$ model exhibits similar characteristics; although the $\Gamma$ and $Z$ pockets shift upward, they still cross the Fermi energy [see Fig.~\ref{Fig4_Tc}(b) for DFT and DMFT Fermi surface and the enhanced occupation in Ni-3$d_{x^2-y^2}$], suggesting that correlations suppress the self-doping effect between Ni-3$d$ and La-5$d$ [these pockets are not part of the  Ni-3$d$ (only) model]. The calculated $d_{x^2-y^2}$ effective electron masses $m^*/m=\frac{1}{Z}$=$1 - \left.\frac{\partial \mathrm{Im} \Sigma(i\omega_n)}{\partial i\omega_n}\right|_{i\omega_n \to 0}$ are with 2.8 (Ni-3$d$) and 3.1 (La-4$f$-5$d$+Ni-3$d$) quite similar for the two models, confirming that the Ni-3$d$ orbital retains its distinct electronic character in both models and is largely decoupled from other orbital contributions.

\emph{Superconductivity in La$_2$NiO$_4$:La$_2$AlO$_4$---}Going even beyond the DMFT correlations,  we compute the superconducting critical temperature ($T_{\mathrm{c}}$) of La$_2$NiO$_4$:La$_2$AlO$_4$ using three different methods: D$\Gamma$A, FLEX, and DCA. For these more involved calculations we use a single-band tight-binding Hamiltonian obtained from a Wannier projection onto the only correlated band crossing the Fermi energy: the Ni-3$d_{x^2-y^2}$ band. Computational details are provided in Supplemental Material Secs.~IV–VIII \cite{SM}. 
As summarized in Table~\ref{Tab1}, the DFT+DMFT calculations (La-4$f$+La-5$d$+Ni-3$d$ model) yield an electron occupancy of $n$=0.87 for the  Ni-3$d_{x^2-y^2}$ orbital. This filling was used in the D$\Gamma$A, FLEX, and DCA calculations
to mimic the effect of the pockets and is consistent with the optimal carrier concentration ($\sim$0.85) predicted in a previous study \cite{kitatani2020nickelate}. 
For FLEX and D$\Gamma$A, we also examined additional fillings ($n$=0.80 and 0.90 for FLEX and $n$=0.80 and 0.95 for D$\Gamma$A). 
Figure~\ref{Fig4_Tc}(a) shows the temperature dependence of the superconducting eigenvalues $\lambda_{\mathrm{SC}}$ as a function of $T$ for different $n$.
D$\Gamma$A yields  $T_{\mathrm{c}}$ ($\lambda_{\mathrm{SC}}$=1) values of 53 and 62\,K for $n$=0.87  and  $n$=0.95, respectively, higher than current IL nickelates \cite{lee2023linear,chow2025bulk,kitatani2020nickelate}. This high  $T_{\mathrm{c}}$ can be traced back to the reduced in-plane lattice constant (3.76\,\AA) which weakens correlations ($U$$\approx$5$t$\,--\,6$t$ instead of 7$t$\,--\,8$t$ for LaNiO$_2$ \cite{PhysRevB.100.205138,PhysRevB.101.075107,PhysRevLett.130.166002}) and  makes La$_2$NiO$_4$:La$_2$AlO$_4$  analogous to 4$d$ palladates \cite{PhysRevLett.130.166002}.

The same tendency is observed in  FLEX with even a  $T_{\mathrm{c}}$  of 103\,K and 115\,K at $n$=0.87 and $n$=0.90, respectively; and DCA with cutting-edge $N_c$=12 cluster sites predicts a similar $T_{\mathrm{c}}$ of 127\,K. Taking all three many-body approaches together and even with the most conservative  D$\Gamma$A $T_{\mathrm{c}}$, we can conclude with quite some confidence that La$_2$NiO$_4$:La$_2$AlO$_4$ is a $d$-wave superconductor with a high $T_c$, which we predict to exceed that of IL nickelates. It should be noted that, during the synthesis of the proposed heterostructures, inversion symmetry (centrosymmetry) breaking may introduce additional bands near the Fermi level into the minimal low-energy model, potentially leading to multiband superconducting pairing \cite{LaNiC2EPJB}. In addition, we discussed the effects of lattice, structure, and strain on the electronic structure of the heterostructures \cite{QuantumwellsPRM} in Supplemental Material Sec.~XIV \cite{SM}.

Finally, the Fermi surfaces obtained from the many-body DMFT, FLEX and D$\Gamma$A methods exhibit remarkable similarity to previous results for states where superconductivity and a well-defined holelike Fermi surface emerges \cite{PhysRevLett.130.166002}. Specifically, Figure~\ref{Fig4_Tc}(c,d) show a characteristic shape with pronounced strong momentum-dependent damping at the many-body level instead of 
the  noninteracting DFT Fermi surface in Fig.~\ref{Fig4_Tc}(b). Noteworthily, electronic correlations induce a Lifshitz transition from an electron Fermi surface in DFT to a holelike Fermi surface.  
Our model parameters ($t^\prime$/$t$=-0.20 and $t^{\prime\prime}$/$t$=0.10) of La$_2$NiO$_4$:La$_2$AlO$_4$ indeed fall within the optimal regime identified in the previous study ($t^\prime$/$t$=-0.22, $t^{\prime\prime}$/$t$=0.14) \cite{PhysRevLett.130.166002}, 
suggesting that this system possesses strong potential for high-$T_{\mathrm{c}}$ superconductivity.

\emph{Conclusions---}Lanthanum nickelates have recently reemerged at the focus of condensed matter research, following the discovery of unconventional cupratelike superconductivity. However, electron doping in nickelates remains much more difficult than in cuprates due to  the highly ionic nature of Ni and limited availability of suitable elements. In contrast, in cuprates, the low ionization energy of their $A$-site elements allows Ce to adopt a 4+ valence state,  enabling electron doping.

In this work, we proposed  intercalation of the insulating RP nickelates as a promising route to achieve electron doping. Electron doping was confirmed using DFT and DMFT calculations: In La$_2$NiO$_4$:La$_2$AlO$_4$, a 3$d^{9–\delta}$ ($\delta\sim0.13$) valence state was realized, demonstrating that the Ni $e_g$ orbital in La$_2$NiO$_4$ is  effectively doped with electron(s) from the additional (LaO)$^+$ layer(s).
We also demonstrated the feasibility of this approach for La$_3$Ni$_2$O$_7$. 
Let us emphasize that the approach is not at all restricted to nickelates. It is widely applicable to introducing electron doping in other RP-phase oxides
and very promising for, e.g., cuprates such as La$_2$CuO$_4$ and ruthenates 
such as Sr$_2$RuO$_4$.

Subsequent many-body calculations, based on a simplified Ni-3$d_{x^2-y^2}$ model including correlation effects for the doped La$_2$NiO$_4$ layers, reveal that the electron-doped compounds host high-$T_{\mathrm{c}}$ $d_{x^2-y^2}$-wave  superconductivity, as shown by D$\Gamma$A, DCA, and FLEX. 
Compared to IL nickelate superconductors, La$_2$NiO$_4$:La$_2$AlO$_4$ exhibits a higher $T_{\mathrm{c}}$ due to the smaller in-plane lattice constant [that is close to 3.868\,\AA~of LSAT: (LaAlO$_3$)$_{0.3}$(Sr$_2$TaAlO$_6$)$_{0.7}$].
This in turn reduces correlations and (antiferromagnetic) spin fluctuations and, thus, enahnces $T_c$. Furthermore, due to the large La pockets, La$_2$NiO$_4$:La$_2$AlO$_4$ naturally resides in the optimal doping regime of superconductivity in the two-dimensional Hubbard model even without any additional doping.

In parallel to the charge-transfer mechanism proposed here, recent work \cite{QuantumwellsPRM} has demonstrated that geometrical effects---such as varying the thickness ratio of quantum wells in superlattices---can serve as an effective tuning parameter for controlling multiband superconductivity and achieving a multigap superconducting state. These findings suggest that a combination of charge transfer (chemical doping) and geometrical engineering could provide even greater flexibility in designing and optimizing the superconducting state in RP-phase nickelate heterostructures.

\emph{Acknowledgments---}L.~S.~acknowledges support from the National Natural Science Foundation of China (Grant No.~12422407) and 
the Key Research and Development Program of Shaanxi (2024QY2-GJHX-42).
K.~H.~and L.~S.~acknowledge funding through the Austrian Science Funds (FWF) Project Grant DOI 10.55776/I5398 and the European Research Council through ERC-2024-ADG RealSuper project DOI 10.3030/101201037.
W.-F.~W.~acknowledges the funding from the China Scholarship Council (CSC).
G.~J.~and M.~J.~acknowledge the support of the National Natural Science Foundation of China (Grant No.12174278).
N.~W.~acknowledges support
through the Deutsche Forschungsgemeinschaft (DFG) funded SFB 1170 (``Tocotronics'', project
No.~258499086). N.~W.~and K.~H.~acknowledge support from the DFG research unit
FOR5249 (``QUAST'', project No.~449872909) and the FWF-funded QUAST (sub)project DOI 10.55776/ KIN2563725.
M.~K.~acknowledges support from JSPS KAKENHI Grant No.~JP23H03817, No.~JP24K17014, and No.~JP25K00961.
Calculations have been done on the National Supercomputing Center in Northwest University (Xi'an), the Vienna Scientific Cluster (VSC), Soochow University, and the Research Institute for Information Technology, Kyushu University (General Projects).

This project is funded in part by the European Union. Views and opinions expressed are however those of the author(s) only and do not necessarily reflect those of the European Union or the European Research Council Executive Agency. Neither the European Union nor the granting authority can be held responsible for them.

\emph{Data availability---}The data that support the findings of this article are openly available \cite{data}.



\clearpage
\onecolumngrid
\setcounter{section}{0}
\setcounter{figure}{0}
\setcounter{table}{0}
\setcounter{equation}{0}
\renewcommand{\thefigure}{S.\arabic{figure}}
\renewcommand{\theequation}{S.\arabic{equation}}
\renewcommand{\thesection}{S.\arabic{section}}
\renewcommand{\thetable}{S.\Roman{table}}

\section*{Supplementary materials to ``Heterostructuring as Gateway to Electron Doping of Nickelate Superconductors''}

\section{I.~Density-functional theory (DFT)}

DFT-level structural relaxations and magnetic and electronic band structure calculations were performed using the projector-augmented-wave (PAW) method \cite{PhysRevB.50.17953} in \verb|Vasp| \cite{kresse1996efficiency,PhysRevB.54.11169} and the  (linearized) augmented plane-wave and local-orbitals [FP-(L)APW+lo] basis set in \verb|Wien2K| \cite{blaha2001wien2k,Schwarz2002} with the Perdew-Burke-Ernzerhof version of the generalized gradient approximation (GGA-PBE) \cite{PhysRevLett.77.3865}. A dense $k$-mesh of 13$\times$13$\times$5 for the unit cell of La$_2$NiO$_4$:La$_2$$X$O$_4$  is adopted for sampling the Brillouin zone. The interaction parameters for Ni-3$d$ and La-5$d$ orbitals were obtained from the constrained random phase approximation (cRPA) \cite{PhysRevB.77.085122} 
(implemented in \textsc{VASP} \cite{PhysRevMaterials.9.015001}) for closely related infinite-layer nickelates before\cite{PhysRevLett.124.166402}. 

All input and output  parameters (and files)  of our DFT calculations as well as of the other calculations detailed below can be found in the data repository Novel Materials Discovery (NOMAD) repository,
10.17172/NOMAD/2026.01.26-2, published together with this manuscript.

\section{II. ~Self-consistent Phonon (SCPH) calculation and \emph{ab initio} Molecular dynamic (AIMD)}

We employ a combination of the supercell finite-displacement method to determine both harmonic and anharmonic interatomic force constants (IFCs). Here, the harmonic IFCs are obtained by least-squares fitting to finite-displacement data, while anharmonic IFCs were extracted from $ab$ $initio$ molecular dynamics (AIMD) trajectories at 300\,K, using a training set generated by random atomic perturbations. Sparse anharmonic potential energy surfaces including terms up to fourth order are constructed using the LASSO method. IFCs of different orders are obtained by applying appropriate real-space cutoff radii. All calculations are performed with high-precision DFT calculations using \verb|VASP| code \cite{kresse1996efficiency,PhysRevB.54.11169}, together with the \verb|Phonopy| \cite{togo2015first} and \verb|ALAMODE| packages \cite{alamodePRL,alamodePRB}.

In main text, we investigate the dynamical and thermodynamic stability of La$_2$NiO$_4$:La$_2$AlO$_4$. 
The phonon spectra are calculated at the DFT level using  \verb|VASP| \cite{PhysRevB.54.11169,kresse1996efficiency}, where harmonic and anharmonic interatomic force constants were obtained by combining the supercell finite-displacement method implemented in the \verb|Phonopy| code \cite{togo2015first} with \textit{ab initio} molecular dynamics, and subsequently evaluated using the \verb|ALAMODE| package \cite{alamodePRL,alamodePRB}. 
On-the-fly machine-learned force-field (MLFF) molecular dynamics simulations \cite{PhysRevLett.122.225701,PhysRevB.100.014105} were performed within \verb|VASP| \cite{PhysRevB.48.13115} for a 2$\times$2$\times$1 supercell of La$_2$NiO$_4$:La$_2$AlO$_4$ containing 64 atoms. 

For further studying effects of
anharmonicity we further use the self-consistent phonon (SCPH) method, which renormalizes phonon frequencies through a self-consistent inclusion of higher-order force constants. In this approach, an effective dynamical matrix is constructed by incorporating quartic anharmonic contributions that depend explicitly on phonon eigenvectors and thermal occupation numbers. At each iteration, the renormalized dynamical matrix is diagonalized to obtain updated phonon frequencies and eigenvectors, which are then used to compute the phonon covariance matrix. This matrix includes both quantum zero-point motion and finite-temperature effects through the Bose-Einstein distribution. To improve numerical stability, a linear mixing scheme is applied to the covariance matrix during the self-consistent cycle. The procedure is iterated until convergence of the phonon frequencies is achieved.

The converged renormalized phonon modes define an effective anharmonic dynamical matrix, from which temperature-dependent phonon dispersions are obtained. In the classical limit, the phonon occupation factors reduce to their high-temperature forms, simplifying the covariance matrix. Finally, the difference between the self-consistently renormalized and harmonic dynamical matrices is Fourier-transformed back to real space to yield anharmonic corrections to the IFCs, which can be used in subsequent lattice-dynamical calculations. Further theoretical details can be found in Refs. \cite{alamodePRL,alamodePRB}.

\section{III.~Dynamical mean-field theory (DMFT)}
Starting points for the DMFT calculations are the \verb|Wien2K| bands projected onto  maximally localized Wannier functions \cite{PhysRev.52.191,RevModPhys.84.1419} using the \verb|Wien2Wannier| \cite{mostofi2008wannier90,kunevs2010wien2wannier} interface for:
(i)  a multi-band model including the full Ni-3$d$,
(ii) on top of this also including the La 5$d$ and 4$f$ orbitals, and
(iii) a single Ni-3$d_{x^2-y^2}$ model with filling adjusted to (ii).
 The thus obtained DFT-Wannier Hamiltonian is supplemented by a local Kanamori interaction, and we employ the fully localized limit as the double counting correction \cite{PhysRevB.52.R5467,PhysRevB.48.16929}.
The cRPA interaction parameters for multi-band DMFT (and spin-polarized DFT+$U$ calculations for magnetic energies) are $U$(Ni-3$d$)=4.40\,eV, $J$(Ni-3$d$)=0.65\,eV and $U$(La-5$d$)=2.50\,eV, $J$(La-5$d$)=0.25\,eV, with $U'$=$U$-2$J$=3.10\,eV and 2.00\,eV for Ni-3$d$ and La-5$d$ orbitals, respectively. The resulting many-body Hamiltonian is solved at room temperature (300\,K) within DMFT employing the continuous-time quantum Monte Carlo solver in the hybridization expansions \cite{RevModPhys.83.349} of the \verb|W2dynamics| code \cite{PhysRevB.86.155158,wallerberger2019w2dynamics}. 
Real-frequency spectra are obtained  via analytic continuation using the maximum entropy method (MaxEnt) \cite{PhysRevB.44.6011,PhysRevB.57.10287} as implemented in the \verb|ana_cont| code \cite{kaufmann2023ana_cont}.

\section{IV. Simplified 3$d_{x ^2-y ^2}$ one-band Hamiltonian}
For the three many body methods beyond DMFT that we discuss below, we
 construct a (simplified) two-dimensional single-band Hubbard model with the properly
translated doping according to the Hamiltonian of: 

\begin{align}
    {\cal H}= \sum_{\mathbf{k},\sigma}
    \epsilon_{\mathbf{k}} c^{\dag}_{\mathbf{k},\sigma} c_{\mathbf{k},\sigma}
    +U \sum_{i} n_{i,\uparrow}n_{i,\downarrow},
\end{align}
where $c^{\dag}_{\mathbf{k},\sigma}(c_{\mathbf{k},\sigma})$ is the creation (annihilation) operator,
\begin{align}
    \epsilon_{\mathbf{k}}=
    -2t[{\rm cos}(k_x)+{\rm cos}(k_y)]
    - 4t^{\prime}{\rm cos}(k_x){\rm cos}(k_y) \nonumber -2t^{\prime\prime}[{\rm cos}(2k_x)+{\rm cos}(2k_y)],
\end{align}
the energy-momentum dispersion, and $U$ the onsite Coulomb repulsion, and $t$=-431\,meV, $t^{\prime}$=87\,meV, $t^{\prime\prime}$=-46\,meV, respectively, are the nearest, second nearest, and third nearest hoppings of Ni-3$d_{x^2-y^2}$ orbital along $R$=(100), (110), and (200). These hoppings lead to $t'$/$t$=-0.20 and $t''$/$t$=0.11. The model parameters are obtained from density-functional theory (DFT),
using  \textsc{wien2k}  \cite{blaha2001wien2k,Schwarz2002} with the PBE \cite{PhysRevLett.77.3865}
exchange correlation functional and  \textsc{wien2wannier} \cite{kunevs2010wien2wannier} for projecting onto a maximally localized Ni-3$d_{x^2-y^2}$ Wannier orbital \cite{RevModPhys.84.1419}.

The (simplified) tight-binding Hamiltonian is constructed with $t$=-431\,meV, $t'$=87\,meV and $t''$=-46\,meV, leading to $t'$/$t$=-0.20 and $t''$/$t$=0.10. The single-band interaction Coulomb $U$ is determined as $U$=5.25$t$=2.25\,eV. Nevertheless, as discussed in our earlier work \cite{kitatani2020nickelate}, the omission of the frequency-dependent nature of $U$ in our current calculations likely leads to an underestimation of its effective strength. We therefore suggest that a slightly larger value, around 
$U$=6$t$=2.58\,eV, provides a more realistic representation of the on-site Coulomb interaction and is hence employed.

To account for self-doping from the La pockets which is beyond this one-band Hamiltonian, we set its filling to that of the many-orbital DFT+DMFT calculations.

\section{V.~dynamical vertex approximation (D$\Gamma$A)}
D$\Gamma$A \cite{Toschi2007,galler2019abinitiodgammaa, held2008dynamical,RevModPhys.90.025003} is employed for including non-local 
correlations beyond DMFT and to calculate  the superconducting $T_c$ of La$_2$NiO$_4$:La$_2$$X$O$_4$ is calculated with  the D$\Gamma$A \cite{galler2019abinitiodgammaa, held2008dynamical,RevModPhys.90.025003}.
The D$\Gamma$A calculations are based on the one-band Hamiltonian from the previous Section. In D$\Gamma$A we first compute the particle-particle vertex incorporating spin fluctuations from both the particle-hole and transversal particle-hole channels. Next, we extract the leading superconducting eigenvalues in the particle-particle channel \cite{PhysRevB.99.041115}. This procedure effectively represents the first iteration of the particle-particle channel in a full parquet D$\Gamma$A calculation \cite{RevModPhys.90.025003}. For further technical details, see \cite{Kitatani2022}.

\section{VI.~Fluctuation exchange (FLEX)}

To investigate the potential for superconductivity and estimate  the critical temperature ($T_c$) in La$_2$NiO$_4$:La$_2$$X$O$_4$ ($X$=Al,\,Ga,\,Sc) via the fluctuation exchange (FLEX) approximation \cite{PhysRevLett.62.961,PhysRevB.43.8044}, we analyze their electronic structures using the model Hamiltonians from Section~IV, which is derived from first-principles calculations.

The FLEX approximation, a self-consistent and conserving approach for strongly correlated systems, is then employed to evaluate many-body effects and superconducting properties. FLEX accounts for dynamic spin and charge fluctuations by performing infinite summations of ring and ladder diagrams in both particle–hole and particle–particle channels. The method determines the electron self-energy through a self-consistent Dyson equation, yielding renormalized Green’s functions and corresponding dynamic susceptibilities. These susceptibilities form the basis for constructing the effective pairing interaction, which is inserted into the linearized Eliashberg equation to identify superconducting instabilities and dominant pairing symmetries. 
For the imaginary-time and Matsubara frequency grids we applied the sparse-sampling approach \cite{PhysRevB.101.035144,PhysRevB.103.205148} in combination with the intermediate representation (IR) basis \cite{PhysRevB.96.035147,wallerberger2023sparse}.
Further methodological details can be found in 
Ref.~\cite{PhysRevB.103.205148,10.21468/SciPostPhys.15.5.197} where similar transition metal oxides are studied.

\section{VII.~Dynamical Cluster Approximation (DCA)}
\label{SMSecDCA}

Again we use the single-band square lattice Hamiltonian from Section~IV

in our  DCA calculations with a continuous-time auxiliary field QMC (CT-AUX) cluster solver \cite{hettler1998nonlocal,maier2005quantum,maier2006pairing,gull2008continuous}. The DCA captures short-range correlations accurately by solving a cluster problem embedded in a self-consistent host, which approximates long-range physics in the mean-field level. With this finite cluster approximation, the first Brillouin zone is divided into $N_c$ patches $\mathcal{P}_{\mathbf{K}}$, each of which is represented by a cluster momentum $\mathbf{K}$. Then, the continuous self-energy $\Sigma_c(\mathbf{k}, i\omega_n)$ is approximated by a piecewise constant function $\Sigma_c(\mathbf{K}, i\omega_n)$ with discrete values on the $N_c$ patches in the momentum space. A coarse-grained average of the momentum $k$ over the entire Brillouin zone yields the averaged Green function. 
\begin{align*}
\bar{G}(\mathbf{K}, i\omega_{n}) = \frac{N_c}{N} \sum_{\mathbf{k} \in \mathcal{P}_{\mathbf{K}}} [i\omega_{n} + \mu - \varepsilon_{\mathbf{k}} - \Sigma_c(\mathbf{K}, i\omega_{n})]^{-1}.
\end{align*}
The cluster Green function and self-energy can be calculated via various quantum many-body methods such as Quantum Monte Carlo (QMC) simulations. 
However, to avoid overcounting  interaction 
and self-energy,a cluster-self-energy-excluded Green function defined as
\begin{align*}
\mathcal{G}_{0}(\mathbf{K}, i\omega_{n}) = [\bar{G}^{-1}(\mathbf{K}, i\omega_{n}) + \Sigma_{c}(\mathbf{K}, i\omega_{n})]^{-1} \, .
\end{align*}
With this $\mathcal{G}_{0}$ and $U$ a cluster is solved to yield a interacting Green's function $G(\mathbf{K}, i\omega_{n})$ and self-energy $\Sigma(\mathbf{K}, i\omega_{n})$ in  self-consistent loop.  

\section{VIII. Estimating $T_c$ from the Bethe-Salpeter Equation}
To estimate $T_c$, we have solved, for all three many-body methods, the Bethe-Salpeter equation (BSE) in the particle-particle channel \cite{maier2006structure}:

\begin{align*}
\lambda_{\nu}(T) \phi_{\nu}(k) = -\frac{T}{N} \sum_{k^{\prime}} \Gamma(k, k^{\prime}) G(k^{\prime}) G(-k^{\prime}) \phi_{\nu}(k^{\prime}).
\end{align*}

Here, $k$ = ($k_x$, $k_y$, $k_z$, $\omega_n$) with $\omega_n$=(2n+1)$\pi$$T$ the fermionic Matsubara frequency and $G$($k$) the fully dressed single-particle propagator. $\Gamma(k, k^{\prime})$ is the particle-particle irreducible vertex function. The system reaches the superconducting transition at $T$=$T_c$ when the leading eigenvalue $\lambda(T_c)$=1. 

\section{IX.~Molecular dynamics simulations results}
\label{SMSec:MD}

\begin{figure*}[ht]
\centering
\includegraphics[width=0.8\linewidth]{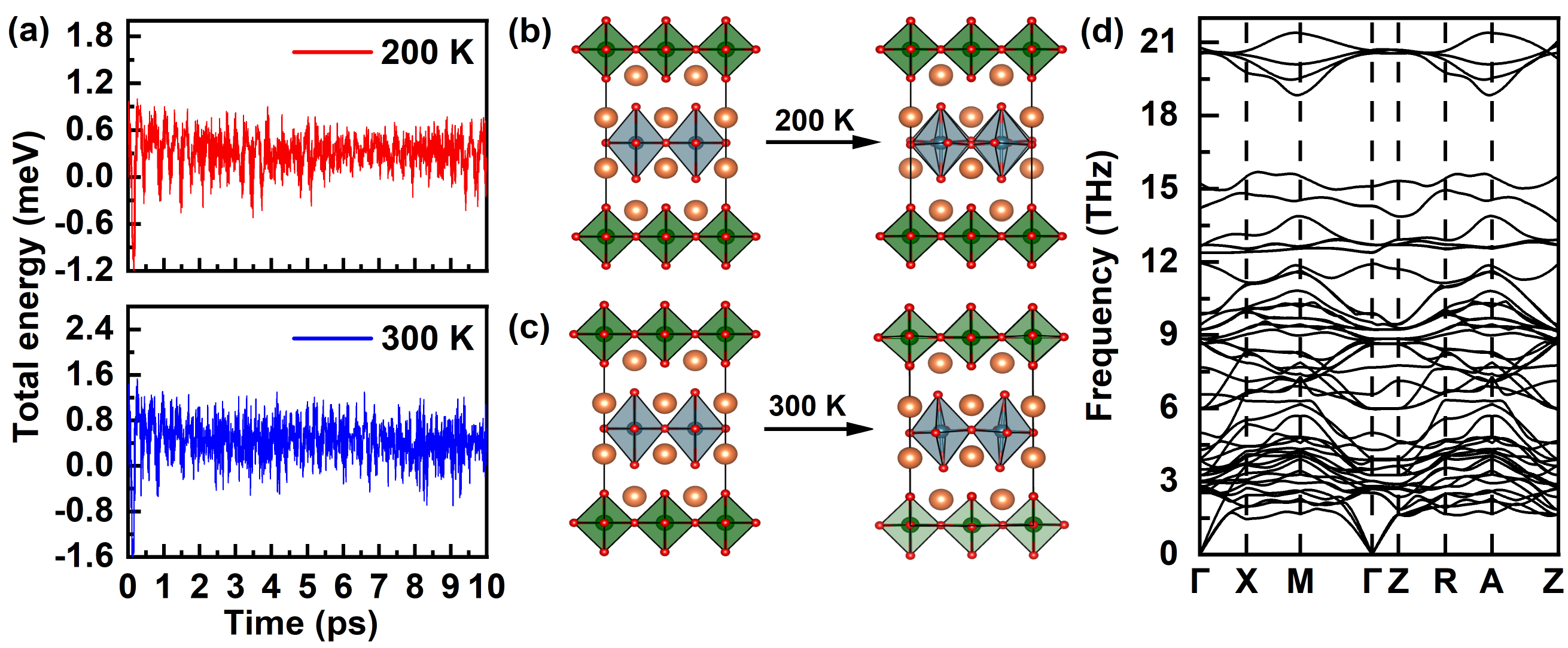}
\caption{(a-c) The molecular dynamics simulation total energy for per La$_2$NiO$_4$:La$_2$AlO$_4$ cell at 200\,K and 300\,K, respectively. (d) Finite-temperature (300\,K) phonon dispersions calculated using the self-consistent phonon (SCPH) method.
}
\label{FigS2_MD}
\end{figure*}

\emph{Structural stability---}Next, we investigate the dynamical and thermodynamic stability of La$_2$NiO$_4$:La$_2$AlO$_4$. 
The phonon calculations and the on-the-fly
machine-learned force-field (MLFF) molecular dynamics \cite{PhysRevLett.122.225701,PhysRevB.100.014105} are performed at the DFT level using  \verb|VASP| \cite{PhysRevB.54.11169,kresse1996efficiency}, where harmonic and anharmonic interatomic force constants were obtained by combining the supercell finite-displacement method implemented in the \verb|Phonopy| code \cite{togo2015first} with AIMD \cite{PhysRevB.48.13115}, and subsequently evaluated using the \verb|ALAMODE| package \cite{alamodePRL,alamodePRB}.
Figure~\ref{FigS2_MD}(a-c) show the total-energy and structural evolution during MLFF molecular dynamics simulations from 300\,K to 50\,K. After around 200-300 steps (corresponding to 0.2-0.3\,ps), only minor thermally induced fluctuations were observed within 2\,meV/unit cell, with no bond-breaking or structural phase transitions.
As shown in Fig.~\ref{FigS2_MD}(b), the phonons (at 300\,K) exhibit no imaginary modes, confirming dynamical stability. These results demonstrate both the dynamical and thermodynamic stability of La$_2$NiO$_4$:La$_2$AlO$_4$ even up to 300\,K.

Fig.~\ref{FigS2_MD} shows the time evolution of the energy in the molecular dynamics simulations at different temperatures and the subsequent lattice configurations obtained in the end step of the simulations.

\begin{figure*}[h!]
    \centering    \includegraphics[width=0.8\linewidth]{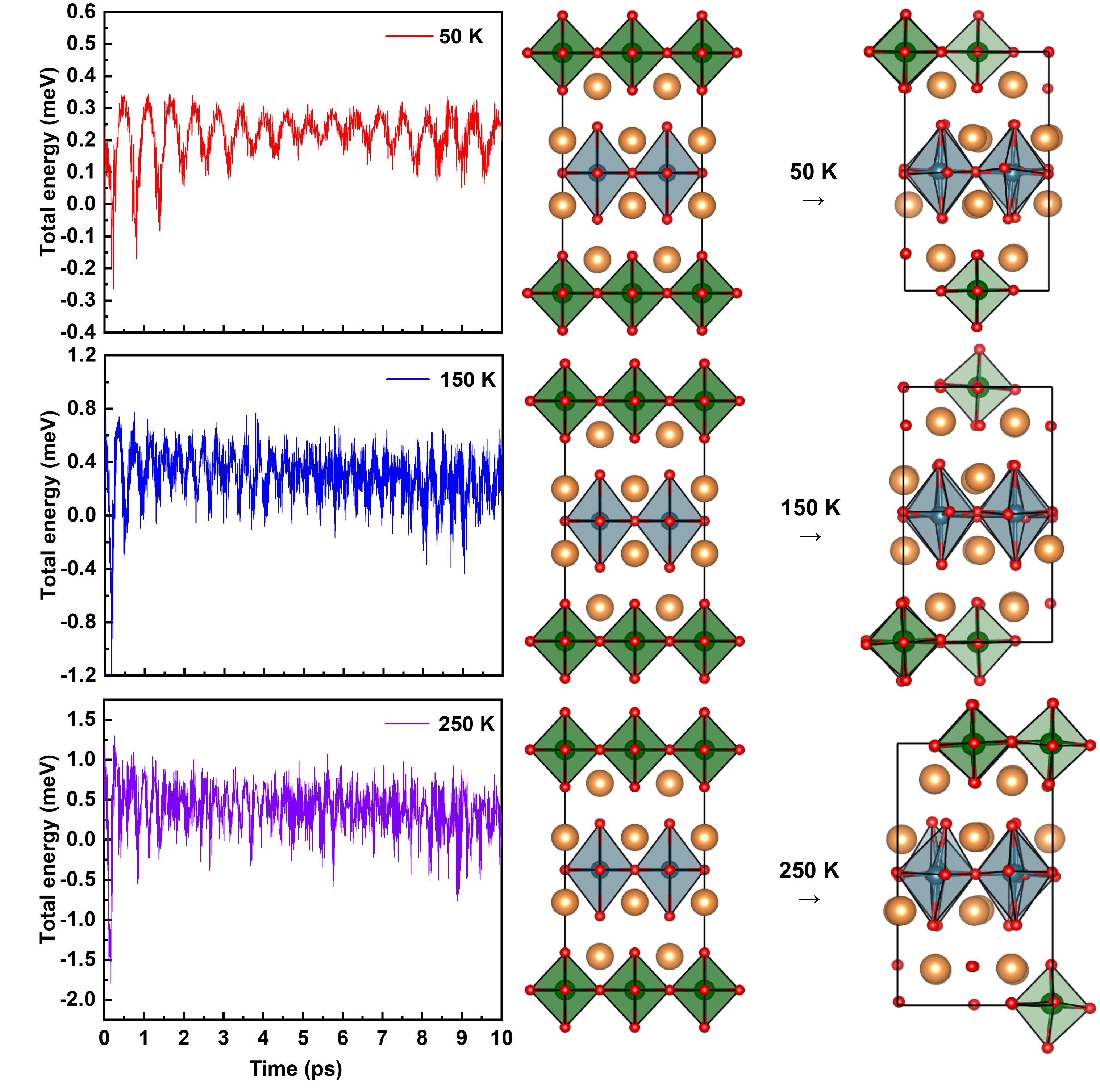}
    \caption{Molecular dynamics simulations results of total energies (in unit of meV/unit cell) for La$_2$NiO$_4$:La$_2$AlO$_4$ at 50\,K, 150\,K and 250\,K.}
    \label{FigS3_MD}
\end{figure*}

\section{X. ~Pseudo-quaternary convex hull}
\begin{figure*}[h!]
    \centering    \includegraphics[width=0.5\linewidth]{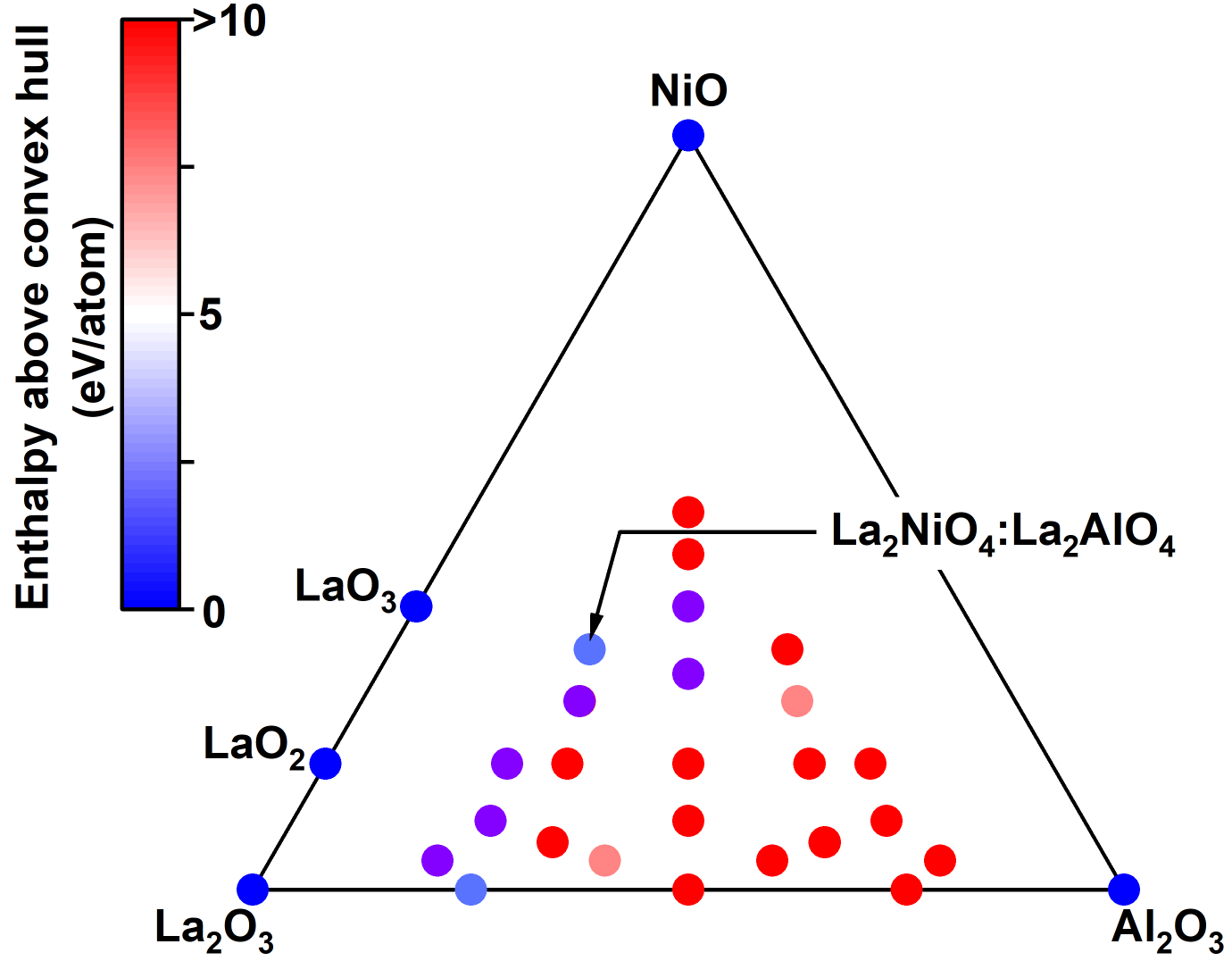}
    \caption{The pseudo-quaternary convex hull of the La-Ni-Al-O system at 0\,K. Stable phases are denoted by bluesolid circles, while the color bar refects the enthalpy (eV/atom) of the compounds above the convex hull.}
    \label{FigS3-convex_hull}
\end{figure*}

We performed a structural search for the La–Ni–Al–O system across a range of compositions. Based on the Inorganic Crystal Structure Database (ICSD), Al$_2$O$_3$, La$_2$O$_3$, and NiO are all thermodynamically stable phases and are widely realized experimentally. Accordingly, we constructed the pseudo-quaternary convex hull using these three oxides as end members. To ensure convergence of the enthalpy within 1\,meV per atom, we employed a $k$-point spacing of 2$\pi$$\times$0.03\,\AA$^{-1}$ and a plane-wave cutoff energy of 800 \,eV for Brillouin zone sampling. The results show that all three binary oxides lie on the convex hull, confirming their stability. The DFT-calculated pseudo-quaternary convex hull for the La$_2$NiO$_4$:La$_2$AlO$_4$ heterostructure is shown, with the proposed structure indicated by label, as shown in Fig.~\ref{FigS3-convex_hull}. Its position and value on the convex hull demonstrates its chemical stability against decomposition into competing phases.

\section{XI.~DFT+$U$ magnetic total energies}

Table~\ref{TabS1} summarizes the relative total energies of different magnetic configurations obtained from spin-polarized DFT+$U$ calculations, with the energy of the $G$-type antiferromagnetic (AFM) state taken as the reference. The results show that the $G$-AFM configuration is the magnetic ground state for all systems.

\begin{table}[h]
\centering
\caption{Magnetic total energies of different systems from spin-polarized DFT+$U$ calculations, the presented results are in  meV/unit cell and the energy of the  $G$-AFM phase is set to zero.}
\label{TabS1}
\begin{tabular}{ccccc}
\hline
\hline
Structure                           & FM      & $A$-AFM    & Stripe-AFM    & $G$-AFM  \\ \hline
La$_2$NiO$_4$                       & 348.5  & 348.9   & 154.4             & 0      \\
La$_2$NiO$_4$:La$_2$AlO$_4$         & 210.0  & /        & 210.1        & 0      \\
La$_2$NiO$_4$:La$_2$GaO$_4$         & 159.1  & /        & 159.0        & 0      \\
La$_2$NiO$_4$:La$_2$ScO$_4$         & 242.2  & /        & 52.9        & 0      \\ \hline
\end{tabular}
\end{table}

\section{XII.~Non-magnetic DFT-derived band structure and orbital characters}
\label{SMSec:addDFT}
\begin{figure*}[h!]
    \centering    \includegraphics[width=0.95\linewidth]{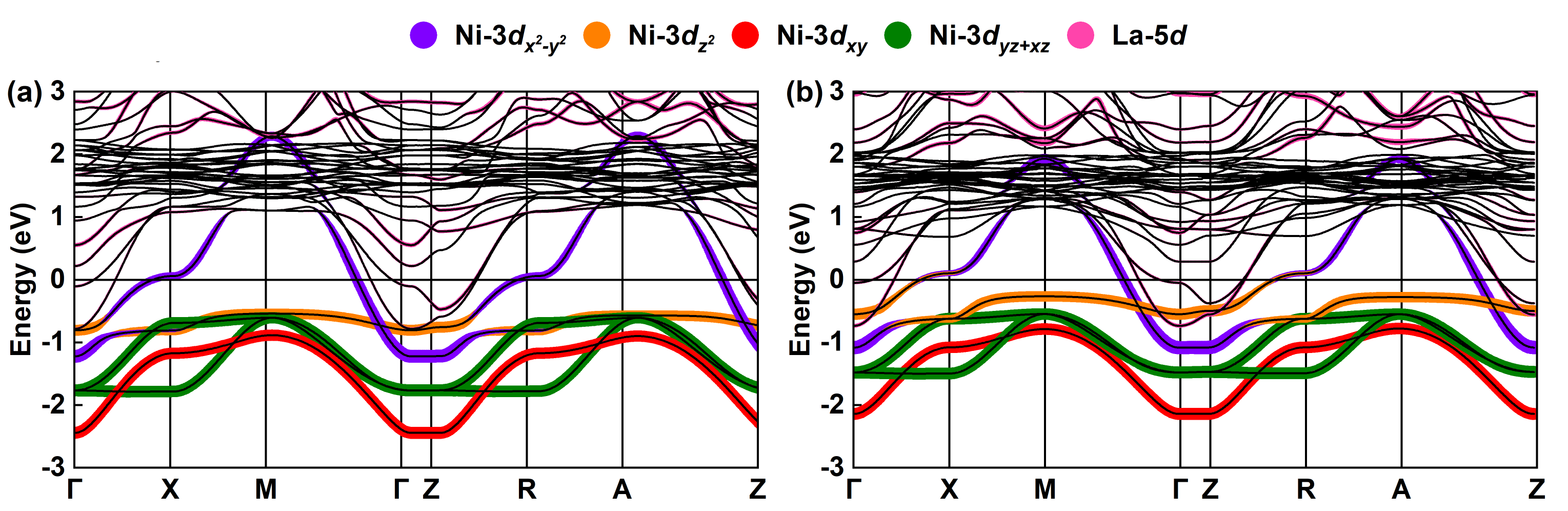}
    \caption{DFT calculated  band structures of (a) La$_2$NiO$_4$:La$_2$GaO$_4$, (b) La$_2$NiO$_4$:La$_2$ScO$_4$.}
    \label{FigS4-band}
\end{figure*}

Supplementing Fig.~2(b)  of the main text, we show in Fig.~\ref{FigS4-band} also the DFT band structure for intercalation with La$_2$GaO$_4$ and La$_2$ScO$_4$.

\section{XIII. ~DFT Charge Transfer analysis}
To further verify the inter-layer charge-transfer effect, we performed both Bader charge analysis and Mulliken population analysis for LaNiO$_2$ (3$d^9$), La$_2$NiO$_4$ (3$d^8$), and the La$_2$NiO$_4$:La$_2$AlO$_4$ heterostructure. As summarized in Table~\ref{tab:charge_analysis}, a net charge transfer from the La$_2$AlO$_4$ layer to the La$_2$NiO$_4$ layer is clearly observed. The resulting Mulliken population and Bader charge on Ni in the La$_2$NiO$_4$:La$_2$AlO$_4$ system is consistent with a Ni$^+$ state and a 3$d^9$ configuration, analogous to that in LaNiO$_2$.

\begin{figure}[tb]
\centering
\includegraphics[width=0.6\linewidth]{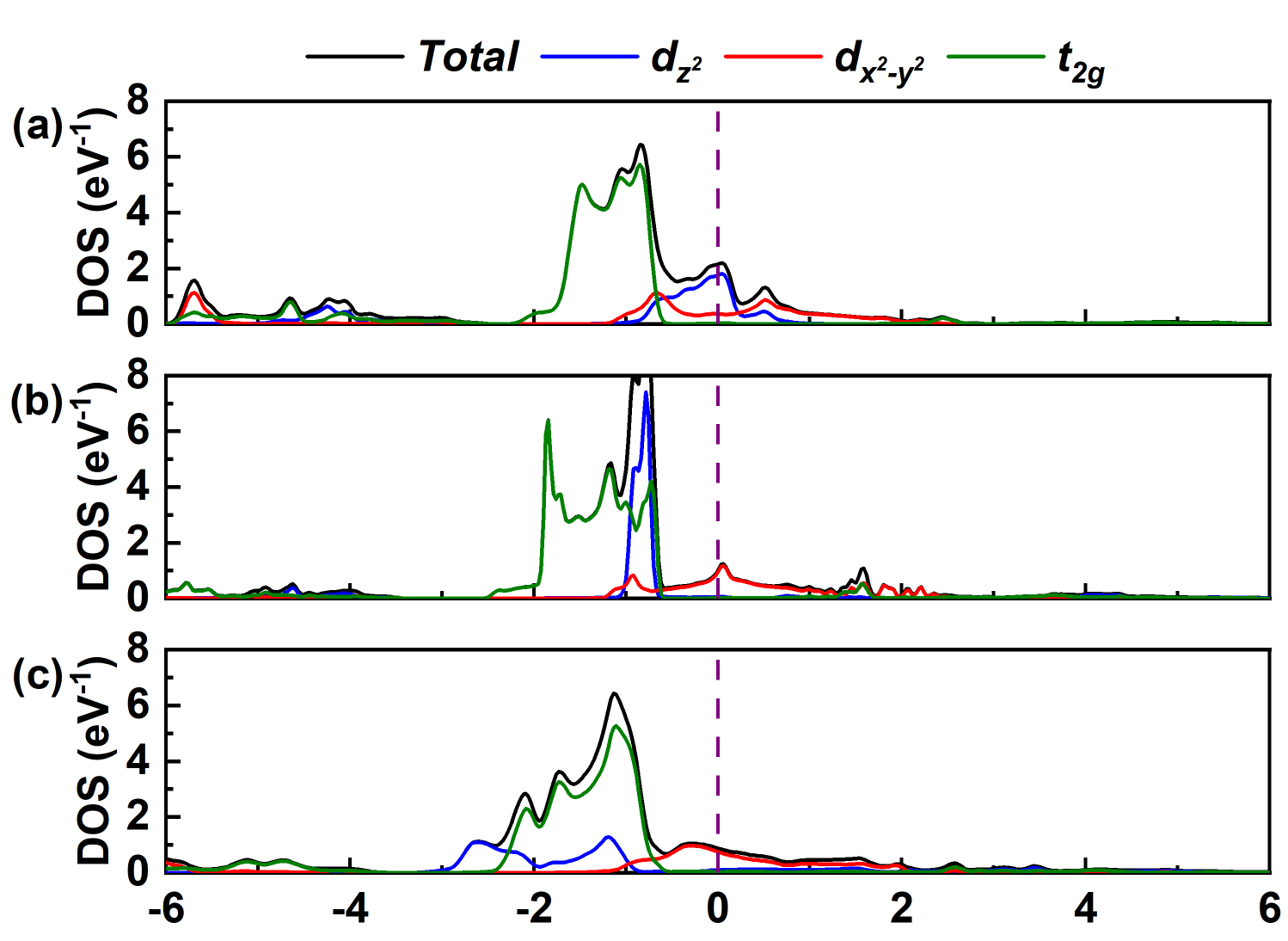}
\caption{DFT density of states (DOS) for (a) La$_2$NiO$_4$, (b) La$_2$NiO$_4$:La$_2$AlO$_4$, and (c) LaNiO$_2$, respectively.}
\label{FigS5-DOS}
\end{figure}

In addition, we further support our conclusions using DFT results by calculating the total and partial density of states for La$_2$NiO$_4$ (Ni$^{2+}$, 3$d^8$), La$_2$NiO$_4$:La$_2$AlO$_4$, and LaNiO$_2$. For La$_2$NiO$_4$ [Fig.~\ref{FigS5-DOS}(a)], the Fermi level is dominated by contributions from the half-filled Ni-3$d_{x^2-y^2}$ and Ni-3$d_{z^2}$ orbitals, consistent with a 3$d^8$ configuration and it features two holes in the Ni-$e_g$ orbital. In contrast, 
for the La$_2$NiO$_4$:La$_2$AlO$_4$ heterostructure, the electronic structure and Fermi surface [Fig.~\ref{FigS5-DOS}(b)] closely resemble those of LaNiO$_2$ (3$d^9$) [Fig.~\ref{FigS5-DOS}(c)], demonstrating that Ni adopts a 3$d^9$ configuration. The above DOS results indicate that the presence of the La$_2$AlO$_4$ layer effectively donates one electron to the La$_2$NiO$_4$ layer.

\begin{table}[htpb]
\centering
\caption{Mulliken population ($Z_{val}$=10) and Bader charge analysis ($Z_{val}$=16) for Ni atom in different phases.}
\label{tab:charge_analysis}
\begin{tabular}{lcccc}
\hline\hline
Phase & Mulliken Population & \multicolumn{1}{c}{Bader Analysis}  \\
   & $d$-occupancy & Total Valence ($e$) \\
\hline
LaNiO$_2$ (Infinite-layer) & 8.70 & 15.33    \\
\hline
 \textbf{La$_2$NiO$_4$:La$_2$AlO$_4$} (Heterstructuring) & \textbf{8.70} & \textbf{15.03}   \\
 \hline
La$_2$NiO$_4$ (Bulk) & 8.57 & 14.88   \\
\hline\hline
\end{tabular}
\end{table}

\section{XIV. ~Quantum Confinement Effect}

In this section, we provide a systematic and comprehensive structural optimization and analysis for a series of related systems, including bulk LaNiO$_3$, LaAlO$_3$, LaScO$_3$, LaGaO$_3$, La$_2$AlO$_4$, La$_2$NiO$_4$, as well as several nickelate/insulator heterostructures such as La$_2$NiO$_4$:La$_2$AlO$_4$, La$_2$NiO$_4$:La$_2$GaO$_4$, La$_2$NiO$_4$:La$_2$ScO$_4$, La$_2$NiO$_4$:
La$_3$Al$_2$O$_7$, and La$_2$NiO$_4$:La$_4$Al$_3$O$_{10}$, as shown in Table~\ref{tab:lattice_parameters}. From these calculations, we extracted the evolution of the lattice parameters of the La$_2$NiO$_4$ nickelate layer as a function of the thickness of the La$_2$AlO$_4$ insulating spacer. The results show that as the thickness of the insulating layer increases, the in-plane lattice constants ($a$ and $b$) and the in-plane Ni--O bonds of La$_2$NiO$_4$ slightly increases from 3.76\,\AA~to 3.78 \,\AA~and from 1.88\,\AA~to 1.89\,\AA, respectively, while  the out-of-plane Ni--O bonds lengths remain almost unchanged from 2.56\,\AA~to 2.57\,\AA. This behavior originates from lattice mismatch and interfacial coupling between the layers; however, the thickness of the insulating LaAlO$_3$-based layers is not expected to provide a powerful handle for continuously tuning the electronic structure. Specifically, replacing Al with Ga or Sc in the spacer layer provides a much more effective route to control the strain, as the significantly larger lattice constants of bulk LaGaO$_3$ (3.90\,\AA) and LaScO$_3$ (4.06\,\AA) lead to a substantial increase in the in-plane Ni--O bond length and a corresponding decrease in out-of-plane distances.

These strain-induced lattice modifications have profound consequences for the electronic structure and superconducting properties. The in-plane lattice parameter directly controls the crystal-field splitting between the Ni-3$d$ orbitals: an increase in the in-plane lattice constant, as seen in the La$_2$NiO$_4$:La$_2$ScO$_4$ spacer case, weakens the overlap between Ni-3$d$ and in-plane O-2$p$ orbitals, reducing the crystal-field splitting and lowering the energy of the Ni-3$d_{x^2-y^2}$ orbital, which may potentially shift the system towards a multi-band description. Along the out-of-plane direction, $c$-axis compression shortens the Ni--O bond length, raising the Ni-3$d_{z^2}$ energy due to stronger repulsion. Crucially, this reduced apical bond length also enhances the hybridization between the Ni-3$d_{z^2}$ orbital and the La-5$d$ orbitals of the spacer layer. This interaction lowers the effective energy of the La-5$d$ states, providing a lower channel for charge redistribution and increasing the charge transfer from Ni-3$d$ to La-5$d$ orbitals. The combined effects of lattice and bond length modifications also shift the position of Ni-3$d$ bands and the Fermi level; therefore, maintaining the optimal hole-doping concentration for high-$T_c$ may require additional strategies, such as partial substitution of La$^{3+}$ with Sr$^{2+}$ in the La$_2$AlO$_4$ spacer layer.

Finally, we have explicitly considered the role of quantum confinement. In our proposed heterostructures, the two-dimensionality of monolayer La$_2$NiO$_4$ combined with the reduced in-plane lattice constant in the La$_2$AlO$_4$ spacer induces significant confinement effects. Specifically, the shorter in-plane lattice raises the Ni-3$d_{z^2}$ energy. Under these conditions, the Ni-3$d_{z^2}$ band becomes nearly flat, shows negligible dispersion along the $c$-axis ($k_z$), and becomes fully occupied. This confinement not only breaks the double degeneracy of $d_{z^2}$ and $d_{x^2-y^2}$ found in bulk LaNiO$_3$, but also suppresses the hybridization between these two orbitals, thereby strengthening the validity of the single $d_{x^2-y^2}$ Hubbard band model. However, for the $n = 2$ bilayer La$_3$Ni$_2$O$_7$ and its heterostructures, this confinement effect is weaker as the Ni-$d_{z^2}$ band retains finite $k_z$ dispersion, making their electronic structures more similar to bulk La$_3$Ni$_2$O$_7$. Consequently, the insulating-spacer-layer-induced lattice effects remain the dominant mechanism for tuning the electronic structure in these thin films. In summary, the La$_2$NiO$_4$:La$_2$AlO$_4$ heterostructure provides an optimal balance of charge transfer, strain, and confinement, hosting an ideal geometry for achieving a high $T_c$ exceeding 60~K in D$\Gamma$A calculations.

\begin{table}[htbp]
\caption{DFT calculated Ni-O bond length, lattice parameters ($a$ and $c$, in unit of\,\AA) for relevant bulk materials and the proposed heterostructures.}
\label{tab:lattice_parameters}
\centering
\begin{tabular}{l|l|c|c|c|c}
\hline\hline
\textbf{System Type} & \textbf{Material}  & \textbf{In-plane Ni-O (\AA)} & \textbf{Out-of-plane Ni-O (\AA)} & \textbf{Lattice $a$ (\AA)} & \textbf{Lattice $c$ (\AA)}  \\ 
\hline
Bulk & $\mathrm{LaNiO}_3$ & 1.92 & 1.92 & 3.83 & 3.83  \\
     & $\mathrm{LaAlO}_3$ & / & / & 3.81 & 3.81  \\
     & $\mathrm{LaGaO}_3$ & / & / & 3.90 & 3.90  \\
     & $\mathrm{LaScO}_3$ & / & / & 4.06 & 4.06  \\
     & $\mathrm{La}_2\mathrm{NiO}_4$ & 1.90 & 2.37 & 3.80 & 12.98  \\
     & $\mathrm{La}_2\mathrm{AlO}_4$ & / & / & 3.78 & 12.67  \\
     & $\mathrm{La}_3\mathrm{Ni}_2\mathrm{O}_{7}$ & 1.92 & 2.19 & 3.83 & 20.30  \\
\hline
Heterostructure & $\mathrm{La}_2\mathrm{NiO}_4 : \mathrm{La}_2\mathrm{AlO}_4$ & 1.88 & 2.56 & 3.76 & 13.29  \\
& $\mathrm{La}_2\mathrm{NiO}_4 : \mathrm{La}_3\mathrm{Al}_2\mathrm{O}_7$ & 1.89 & 2.55 & 3.78 & 17.05  \\
& $\mathrm{La}_2\mathrm{NiO}_4 : \mathrm{La}_4\mathrm{Al}_3\mathrm{O}_{10}$ & 1.89 & 2.57 & 3.78 & 20.85  \\
& $\mathrm{La}_2\mathrm{NiO}_4 : \mathrm{La}_2\mathrm{Ga}\mathrm{O}_{4}$ & 1.90 & 2.50 & 3.80 & 13.15  \\
& $\mathrm{La}_2\mathrm{NiO}_4 : \mathrm{La}_2\mathrm{Sc}\mathrm{O}_{4}$ & 1.97 & 2.44 & 3.94 & 12.97  \\
& $\mathrm{La}_3\mathrm{Ni}_2\mathrm{O}_{7} : \mathrm{La}_3\mathrm{Al}_{2}\mathrm{O}_{7}$ & 1.91 & 1.99 & 3.82 & 20.58  \\
& $\mathrm{La}_3\mathrm{Ni}_2\mathrm{O}_{7} : \mathrm{La}_3\mathrm{Ga}_{2}\mathrm{O}_{7}$ & 1.93 & 2.28 & 3.86 & 20.48  \\
& $\mathrm{La}_3\mathrm{Ni}_2\mathrm{O}_{7} : \mathrm{La}_3\mathrm{Sc}_{2}\mathrm{O}_{7}$ & 1.99 & 1.97 & 3.98 & 20.55  \\
\hline\hline
\end{tabular}
\end{table}


%

\end{document}